\documentclass[letterpaper,twocolumn,english,aps,prl,groupedaddress,superscriptaddress,showpacs,amsfonts]{revtex4}
\usepackage[T1]{fontenc}
\usepackage[latin1]{inputenc}
\usepackage{float}
\usepackage{amsmath}
\usepackage{graphicx}
\usepackage{amssymb}

\makeatletter
\usepackage{babel}
\makeatother
\begin{document}

\title{Double Exchange Model for Magnetic Hexaborides}

\author{Vitor~M.~Pereira}
\affiliation{Department of Physics, Boston University, 590 Commonwealth Avenue,
Boston, Massachusetts 02215, USA}
\affiliation{CFP and Departamento de Física, Faculdade de Ciências Universidade
do Porto 4169-007 Porto, Portugal.}

\author{J.~M.~B.~Lopes~dos~Santos}
\affiliation{CFP and Departamento de Física, Faculdade de Ciências Universidade
do Porto 4169-007 Porto, Portugal.}

\author{Eduardo~V.~Castro}
\affiliation{CFP and Departamento de Física, Faculdade de Ciências Universidade
do Porto 4169-007 Porto, Portugal.}

\author{A.~H.~Castro~Neto}
\affiliation{Department of Physics, Boston University, 590 Commonwealth Avenue,
Boston, Massachusetts 02215, USA}

\date{\today}

\begin{abstract}
A microscopic theory for rare-earth ferromagnetic hexaborides, such as
Eu$_{1-x}$Ca$_x$B$_6$, is proposed on the basis of the double-exchange Hamiltonian. In these systems, the
reduced carrier concentrations place the Fermi level near the mobility
edge, introduced in the spectral density by the disordered spin
background.  
We show that the transport properties such as Hall effect, 
magnetoresitance, frequency dependent conductivity, and DC resistivity 
can be quantitatively described within the model. We also make specific 
predictions for the behavior of the Curie temperature, $T_C$, as a 
function of the plasma frequency, $\omega_p$.
\end{abstract}


\maketitle

The series of compounds
$\textrm{R}_{1-x}\textrm{A}_{x}\textrm{B}_{6}$, 
where A is an alkaline-earth metal such as Ca or Sr, and R a rare-earth
magnetic ion, has recently attracted considerable interest, following
a series of experiments which unveiled intriguing connections between
their magnetic, transport and optical properties. These are cubic
compounds  where a divalent \cite{Fisk:1979} lanthanoid occupies
the central position on a cube, surrounded by eight B$_{6}$ octahedra
at each vertex. Boron atoms make up a rigid cage, held together by
covalent bonds between neighboring B atoms. Our focus will be on
the results known for $\textrm{Eu}_{1-x}\textrm{Ca}_{x}\textrm{B}_{6}$.
EuB$_{6}$ is a ferromagnetic metal, ordering at $T_{C}\approx15\,\textrm{K}$,
and characterized by a quite small effective carrier density, of order
of $10^{-3}$ per unit cell, at high temperatures
\cite{Degiorgi:1997,Sullow:1998,Paschen:2000}. 
Magnetism is found to arise 
from the half-filled $4f$ shell of Eu, whose localized electrons
account for the measured magnetic moment of $7\,\mathrm{\mu_{B}}$
per formula unit \cite{Sullow:1998,Henggeler:1998,Paschen:2000}.
The FM transition temperature is reported to decrease with increasing
Ca content \cite{Paschen:2000,Wigger:2002} and the totally substituted
compound CaB$_{6}$ exhibits no magnetism. 

In this paper we propose a simple model that describes quantitatively the
properties revealed by the experiments done in EuB$_{6}$: 
(i) a precipitous drop in the DC resistivity just below
$T_{C}$, with a change by a factor as high as 50 between $T_{C}$ and the lowest
temperatures \cite{Sullow:1998,Paschen:2000}; (ii) the large
negative magnetoresistance observed near $T_C$; (iii) an increase
in the number of carriers, by a factor of $2\,\textrm{--}\,3$,
upon entering the ordered phase, as evidenced by Hall effect
\cite{Paschen:2000}; (iv) a large blue shift of the plasma
edge, seen also for $T\leq T_{C}$, both in reflectivity, $R(\omega)$,
and polar Kerr rotation \cite{Degiorgi:1997,Broderick:2002,Broderick:2003};
(v) a scaling of the plasma frequency with the magnetization. 
The above listed features constitute a case for the strong coupling of
the magnetization to the transport properties. 

The effects of chemical doping with non-magnetic Ca are also considered, and
the theory explains qualitatively the following experimental findings: 
(1) with doping, $x$, the metallic regime, found in EuB$_{6}$ ($x=0$),
evolves to a semiconducting behavior above $T_{C}$
\cite{Paschen:2000,Wigger:2002}; (2) just below $T_C$ the carrier density
increases by at least two orders of magnitude \cite{Paschen:2000}; (3)
the plasma edge is visibly smeared while the corresponding
resonance in the polar Kerr rotation is greatly attenuated
\cite{Perucchi:2003,Caimi:2004}; 
(4) $\rho(T,H)$ and $\omega_{p}$ display
an exponential dependence in the magnetization \cite{Wigger:2002,Wigger:2003};
(5) there remains a significant and rapid decrease of $\rho(T)$ just
below $T_{C}$, albeit by a smaller factor than in the undoped case.
Our model predicts that the square of the plasma frequency,
$\omega_p^2$, scales linearly with the Curie temperature, $T_C$,
as doping $x$ is varied. Furthermore, we also present a phase 
diagram $T$ versus $x$ which contains two phase transition lines:
paramagnetic to ferromagnetic and metal to insulator. 
 
Theoretical understanding of the microscopic mechanisms responsible
for the transport data is still controversial. Earlier local
density approximation (LDA) calculations 
predicted semi-metallic character with an overlap of conduction and
valence bands 
at the Fermi level on the $X$ point of the Brillouin zone
\cite{Massidda:1996,Rodriguez:2000}. 
This appears to be consistent with the Shubnikov--de Haas and de
Haas--van Alphen measurements \cite{Aronson:1999,Foot1}. Further
density functional theory (DFT)  
calculations showed the existence of a sizeable gap of $\sim1\,\mathrm{eV}$.
These results are corroborated by angle resolved photoemission
spectroscopy (ARPES) for magnetic 
\cite{Denlinger:2002} and non-magnetic hexaborides \cite{Souma:2003}.
Some existing theoretical models assume the semimetal scenario
\cite{Cooley:1997,Wigger:2003,Calderon:2003}, 
that ARPES seems to rule out. 

We present a microscopic theory for EuB$_{6}$, involving a single
conduction band, consistent with the observed properties of these
compounds, and make specific predictions regarding  the phase diagram
for Eu$_{1-x}$Ca$_{x}$B$_{6}$. The presence of defects in the crystalline
environment \cite{Noack:1980,Monnier:2001,Denlinger:2002} is believed 
to create a small carrier concentration in the conduction band. 
These carriers interact with the local Eu magnetic
moments in a cubic lattice, via an \emph{s--f} exchange coupling, leading
to the well known Kondo lattice Hamiltonian:
\begin{equation}
\mathcal{H}_{KL}=-\sum_{\left\langle i,j\right\rangle \sigma
}\left(t_{i,j}c_{i,\sigma}^{\dag}c_{j,\sigma}+\textrm{h.c.}\right)+J\sum_{i}\vec{\mathbf{S}}_{i}\cdot\vec{\mathbf{s}}_{i}\,, 
\label{eq:Hamiltonian_KLM}
\end{equation}
where $\vec{\mathbf{S}}_{i}$ represents the $4f$ Eu spin ($S=7/2$)
operator at the site $i$ and $\vec{\mathbf{s}}_{i}=1/2\sum_{\alpha,\beta}c_{i,\alpha}^{\dag}\vec{\mathbf{\tau}}_{\alpha,\beta}c_{i,\beta}$
is the the electron spin operator. Given
the large value of $S$, the core spins are treated as classical
variables parameterized by the polar and azimuthal angles as $\vec{\mathbf{S}}_{i}=S\left(\sin\theta_{i}\cos\phi{}_{i},\sin\theta_{i}\sin\phi{}_{i},\cos\theta_{i}\right)$.
Based on the current literature, reasonable ranges for the model parameters
are $t_{ij}=t\sim0.1\textrm{--}1\,\textrm{eV}$  and
$JS\sim0.35\textrm{--}0.7\,\textrm{eV}$. 

The carrier concentration in EuB$_{6}$ ($n\sim10^{-3}$) is very small,
and therefore the mean kinetic energy is much smaller than the magnetic one, 
$\left\langle K\right\rangle \sim tn\ll JS$.
In this regime, the low energy physics of Hamiltonian
(\ref{eq:Hamiltonian_KLM}) can 
be obtained by taking the limit of $J\to\infty$ \cite{Pereira:2003}.
The effective Hamiltonian is obtained by projecting out the states
with the electronic spin parallel (for $J>0$) or anti-parallel (for
$J<0$) to the local core spin $\vec{\mathbf{S}}_{i}$. In either case,
the result is nothing more than the double-exchange (DE)
Hamiltonian \cite{Anderson:1955}: \begin{equation}
\mathcal{H}=-\sum_{\left\langle i,j\right\rangle }\left(\widetilde{t}_{i,j}c^{\dag}c_{j}+\textrm{h. c.}\right)\,.\label{eq:Hamiltonian_DE}\end{equation}
The effective hoping amplitude is $\widetilde{t}_{i,j}=t\cos\left(\theta_{i}/2\right)\cos\left(\theta_{j}/2\right)+\sin\left(\theta_{i}/2\right)\sin\left(\theta_{j}/2\right)e^{-i\left(\phi_{i}-\phi_{j}\right)}$.

The Hamiltonian (\ref{eq:Hamiltonian_DE}) describes a very interesting
interplay between the magnetic and electronic degrees of freedom. The
spin texture acts as a \emph{static } disordered background for the
electronic motion \cite{Foot2}. The ground state of 
the problem is obtained by minimizing the kinetic energy with the alignment
of all the spins in the system. Thus, the electronic density of state
(DOS), $N(E,M)$, and the Fermi energy, $E_F(M)$, change with
magnetization, $M$, even if the number of electrons is constant. In
this ferromagnetic state the 
transport properties are tied to the magnetic ones. 
Similar physics can be found in the context of the colossal
magnetoresistance (CMR) manganites \cite{Dagotto_Book:2003}. 
This is indeed an intrinsically disordered problem where the strength
of the non-diagonal 
disorder is determined by $M$. A mobility
edge, $E_{C}(M)$, appears in the spectral density \cite{Economou:1972}
and strongly depends on $M$. In the absence of structural
disorder, when the system is fully magnetized ($M=1$) the mobility edge moves
out of the band. One immediately realizes that in the paramagnetic phase
($M=0$), if $E_{C}(0)$ and $E_{F}(M=0)$ are comparable, 
the onset of ferromagnetism at $T_{C}$
and the concomitant displacement of $E_{C}$ towards the bottom of
the band, have a major effect in the transport properties, by
allowing more states to become delocalized and consequently increasing
the conductivity. This was actually a concept that lingered for some
time in the context of the manganites where the metal--insulator transition
and CMR were assumed to stem from this effect \cite{Varma:1996}.
In that case spin disorder alone cannot 
account for the experimental evidence since $E_{C}(0)$ encloses
less than $0.5$ \% of the states \cite{Li:1997}. In the case of EuB$_{6}$,
however, the small number of localized states is comparable to the
concentration of mobile carriers and thus this effect should
have visible consequences. 

\begin{figure}
\begin{center}\includegraphics[%
  bb=18bp 70bp 718bp 546bp,
  clip,
  scale=0.31]{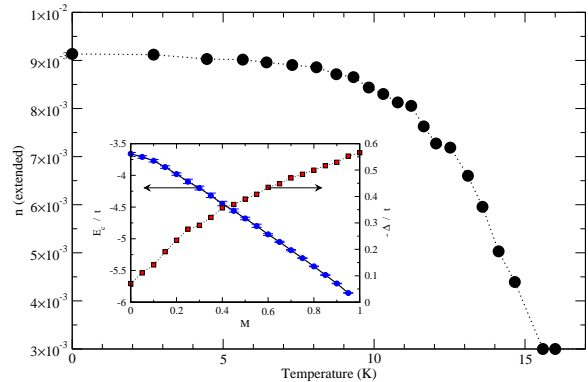}\end{center}

\caption{\label{fig:Density_MobEdge}Evolution of $n_{e}(T)$ assuming
  $n_{e}^{0}=0.003$ 
and $E_{W}=0.1\, t$. The inset shows the mobility edge $\widetilde{E}_{C}$
(circles) and mobility gap $\Delta$ (squares) as functions of the
normalized magnetization $M$.}
\end{figure}

\emph{{}``Pure'' EuB$_{6}$}. In order to investigate the aforementioned,
we calculate $N(E,M)$ and $E_{C}(M)$ for the Hamiltonian
(\ref{eq:Hamiltonian_DE}) 
for static spin configurations with specific values of $M$. 
The full, self-consistent treatment of the magnetism of Hamiltonian
(\ref{eq:Hamiltonian_DE}) is computationally too demanding for the system
sizes required to study localization effects. In this work we 
circumvent these difficulties by assuming uncorrelated spins in an external 
field and extracting the temperature dependence of the zero-field
magnetization, $M(T)$, from the experimental data \cite{Henggeler:1998}.
The recursion method \cite{Haydock:1980}
was used to calculate $N(E,M)$, and $E_{C}(M)$ was located using
the transfer matrix method
\cite{MacKinnon:1981,Pichard:1981}. Non-magnetic defects 
were accounted by a small magnetization independent shift, $E_{W}$,
of $E_{C}$, thus giving effectively $\widetilde{E}_{C}(M)=E_{W}+E_{C}(M)$.
The important quantity
is the mobility gap, $\Delta(M)=\widetilde{E}_{C}(M)-E_{F}(M)<0$,
the variation of which, at the transition, completely determines the
number of extended (metallic) carriers, $n_{e}(M)$. Tuning $E_{F}(0)$
so that $n_{e}^{0}\equiv n_{e}(0)=0.003$, as reported in the transport
measurements above $T_{C}$, the variation of $n_{e}$ is presented in
Fig.~\ref{fig:Density_MobEdge}. $\Delta(M)$ is found to be almost linear 
in $M$: $\Delta(M) = \Delta_0 (1-\alpha M)$ 
(see the inset in Fig.~\ref{fig:Density_MobEdge}). The range of
variation of $n_{e}$ in Fig.~\ref{fig:Density_MobEdge} compares
well with the data obtained from the Hall effect by Paschen \emph{et
al.} \cite{Paschen:2000}.

Another experimentally accessible quantity 
is the plasma frequency, $\omega_{p}$. Applying the Kubo formula for
the optical conductivity, $\sigma(\omega)$, to Hamiltonian
(\ref{eq:Hamiltonian_DE}) the sum rule for its real part, $\sigma'(\omega)$,
can be extracted \cite{Maldague:1977} and combined with the optical
sum rule \cite{Wooten:1972} leading to the plasma frequency for the
model $\omega_{p}^{2}=- 4\pi e^{2}a^{2}\left<\mathcal{H}\right>/3$.

\begin{figure}
\begin{center}\includegraphics[%
  bb=46bp 31bp 706bp 546bp,
  clip,
  scale=0.31]{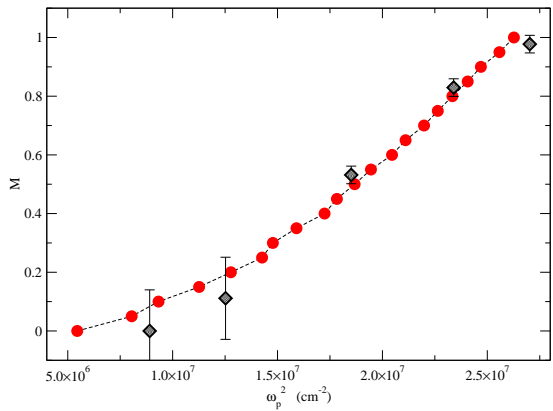}
\end{center}
\caption{\label{fig:Wp}Result for $\omega_{p}^{2}(M)$ obtained from the
sum rule (circles) and experimental data (diamonds). The experimental
($H=0$) points were obtained combining $\omega_{p}^{2}(T)$ at zero
field from ref. \cite{Broderick:2002} with the zero field $M(T)$
from ref. \cite{Henggeler:1998}.}
\end{figure}

In the reported experimental results \cite{Broderick:2002}, $\omega_{p}^{2}$
is obtained from the Drude contribution to the optical response of
the material. Therefore, in order to compare with the model predictions,
we included only the contribution of metallic states in the calculation
of the kinetic energy,
$\omega_{p}^{2}(M)\sim\int_{Ec(M)}^{\infty}N(E,M)f(E)EdE$. The outcome
of such procedure being shown in Fig.~\ref{fig:Wp}.  
The linear relation $\omega_{p}^{2}\sim M$
inferred from ref. \cite{Broderick:2002b}, for high $M$, is indeed
verified. These results for $\omega_{p}$ are obtained with
$t=0.55\,\textrm{eV}$.

The steep drop in the resistivity below $T_{C}$ can be understood
in a consistent way: the resistivity is dominated by spin and phonon
scattering at high temperatures until the magnetic transition is reached.
Below this point the negative mobility gap increases along with $M$,
delocalizing a considerable amount of the previously localized states.
The scaling theory of localization \cite{Lee:1985} prescribes that
$\sigma(M)\sim\left(-\Delta(M)\right)^{\nu}$. Replacing here
the results for $\Delta(M)$ of Fig.~\ref{fig:Density_MobEdge}, we
find that $\rho(T_{C})/\rho(T=0)\sim20\,\textrm{--}\,50$, in agreement
with the values obtained in the experiments \cite{Sullow:1998,Paschen:2000}.
The negative magnetoresistance (MR) comes about as a natural consequence
of our microscopic mechanism and further supports our claim that the
dependence of the mobility gap on the magnetization is the most relevant
factor driving the physics of this material near $T_{C}$. 

\emph{Doped EuB$_{6}$}. Mean-field analysis of Hamiltonian
(\ref{eq:Hamiltonian_DE}) predicts that $T_{C}$ scales with
$\left<\mathcal{H}\right>$ \cite{Gennes:1960,Kubo:1972}.  
Since $\omega_{p}^{2}$ follows the same scaling, as stated above, we
expect that in the series
$\textrm{Eu}_{1-x}\textrm{Ca}_{x}\textrm{B}_{6}$ 
the squared plasma frequency should scale approximately with $T_{C}$,
a prediction that would be interesting to investigate experimentally
with additional infrared reflectivity experiments.

Band structure calculations seem to agree that the conduction band
has a strong $5d$ Eu component. Ca doping not only dilutes the magnetic
system but also the conducting lattice. The hoping parameter
$\widetilde{t}_{i,j}$ 
in eq.~(\ref{eq:Hamiltonian_DE}) is then replaced by
$\widetilde{t}_{i,j}p_{i}p_{j}$, 
where $p_i =1$ if the site $i$ is occupied by a Eu atom and $p_i=0$, otherwise.
The microscopic problem thus becomes a DE problem in a percolating
lattice which, at $T=0\,\textrm{K}$, reduces to a quantum
percolation problem \cite{Shapir:1982}. Since Ca and Eu are isovalent in
hexaborides one does not expect the number of carriers to depend on $x$.
Nevertheless,
since carriers are presumed to arise from defects it is difficult
to be specific on this issue. The mobility edge, on the other hand,
is very sensitive to the Eu$\,\to\,$Ca substitution, and
should drift toward the band center. In the paramagnetic regime ($T>T_{C}$),
$\widetilde{E}_{C}$ should move past the Fermi energy at some critical
doping $x_{MI}^{P}$, after which the mobility gap becomes positive.
This determines a crossover from the metallic regime to an insulating
behavior for $T>T_{C}$, as seen in the doped compounds
\cite{Paschen:2000,Wigger:2002}. 
At finite $T$ the
mobile carriers arise from thermal activation across the mobility
gap. The resistivity should display a semiconducting behavior with
$T$ and its dependence on $M$ should be dominated
by an exponential factor $\rho(M)\sim\exp(\Delta(M)/(k_B T))$ \cite{Mott:1971}.
Using $\Delta(M) \approx \Delta_0(1-\alpha M)$, as happens in the
non-diluted case for either of the $\Delta(M)\lessgtr0$ situations, we
find that $\rho(M) \sim \exp(-\beta M)$ ($\beta$ is a constant), 
as seen in the experiments \cite{Wigger:2002}. 

\begin{figure}[H]
\begin{center}\includegraphics[%
  scale=0.3]{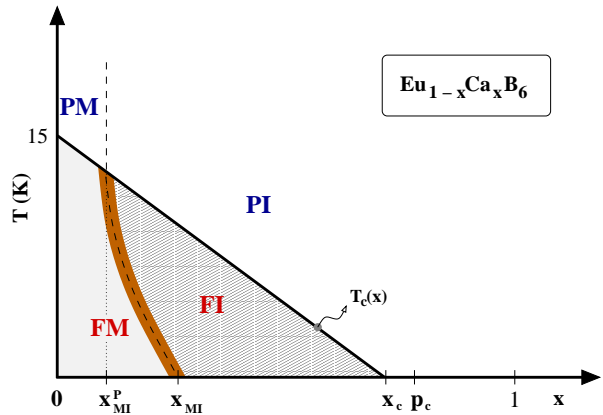}\end{center}

\caption{\label{fig:Phase_Diagram}Schematic phase diagram for
  Eu$_{1-x}$Ca$_{x}$B$_{6}$ 
in the $T\,\textrm{--}\, x$ plane in standard notation: PM stands
for paramagnetic metal, PI for paramagnetic insulator, FM means ferromagnetic
metal and FI ferromagnetic insulator. }
\end{figure}

 As $T$ is lowered below $T_{C}$ and $M$ increases on
the percolating cluster, one expects the crossover from
metallic to semiconducting behavior to occur at larger values of $x$; 
this is illustrated in Fig.~\ref{fig:Phase_Diagram}
by the curved dashed line separating the FM and FI regions. In the
vicinity of this line, a sharp metal--semiconducting distinction is
not possible, resulting possibly in a \emph{bad--metal} behavior.
At $T=0$, there is a metal--insulator transition occurring at a concentration
$x_{MI} \geq x_{MI}^{P}$, which corresponds to the quantum percolation
transition for a small number of carriers (this is different from
the usual quantum percolation point $p_{Q}<p_{c}$, which is defined
by the localization of all states in the band \cite{Chang:1994}).
Even though we expect $x_{MI}$ to be close to $x_{MI}^{P}$, the
possibility of a semiconducting behavior crossing over to metallic
at low $T$  (for some $x_{MI}^{P}<x<x_{MI}$) cannot be excluded.
Ferromagnetism induced by the DE mechanism is expected to persist
past $x_{MI}$ as long as the localization length is greater than
the lattice spacing. Naturally, the critical concentration, $x_{c}$,
where $T_{C}\to0$, should not be higher than $p_{c}\approx0.69$,
the classical site--percolation threshold for the simple cubic lattice
\cite{Stauffer:1994}. The values of $x_{MI}^{P}$ and $x_{MI}$ in
Fig.~\ref{fig:Phase_Diagram} vary with carrier
density and are expected to be sample dependent, since carriers seem to
originate from defects. In fact, annealing experiments can be
quite important for the study of the phase diagram.

Some questions remain as yet unanswered. The value of $t=0.55\,\textrm{eV}$
used in Fig.\ref{fig:Wp} leads to a mean-field
estimate of $T_{C} \approx 80\;\textrm{K}$ \cite{Kubo:1972} which is 
much higher than the experimental value. At such low carrier concentrations, 
magnetic polaron formation \cite{Wegener:2002,Calderon:2003} must certainly
be taken into account close to $T_C$ and could
lead to a reduction of the mean-field estimate. Indeed, 
Raman scattering studies \cite{Nyhus:1997,Snow:2001} close $T_C$ have been
interpreted in terms of magnetic polarons. A large,
positive, MR is observed at high fields and $T<T_{C}$ in EuB$_{6}$
\cite{Aronson:1999,Paschen:2000}. If $J>0$, it is possible that high
magnetic fields may bring into play states of parallel electronic 
and Eu spins, that were projected out to obtain the Hamiltonian
(\ref{eq:Hamiltonian_DE}). One would then have to deal with
two types of carriers and therefore a positive MR. Finally,
our model allows for the unusual situation of a competition
between conduction by thermal activation across a small, magnetization
dependent, mobility gap, and conduction by hopping among localized
states. Some of the details of the transport properties, such as the
remarkable scaling of the Hall coefficient with magnetization in 
Eu$_{0.6}$Ca$_{0.4}$B$_{6}$
\cite{Wigger:2002}, will require a more thorough understanding of
transport in these circumstances.

In conclusion, we have proposed a theory for Eu hexaborides,
based on the close proximity of the Fermi level and a magnetization
dependent mobility edge $E_{C}(M)$. This theory arises quite naturally
from the Kondo lattice  problem in the limit of very small number
of carriers; it accounts, in a quantitative way, for many of the observed
experimental properties in these materials. Furthermore, we also
predict a linear scaling of $\omega_p^2$ with $T_C$ that can be
tested in infrared reflectivity experiments. 

We thank L.~Degiorgi, P.~Littlewood, H.~Ott, and C.~Varma for 
illuminating conversations.
V.M.P., E.V.C. and J.M.B.L.S. are financed by FCT
and the European Union, through POCTI(QCA III).
V.M.P. further acknowledges Boston University for the hospitality and
the financial support of F.C.T., through grant
ref. SFRH/BD/4655/2001. A.H.C.N. was partially supported through
NSF grant DMR-0343790.  

\bibliographystyle{apsrev}
\bibliography{DEModelEuB6}

\begin{thebibliography}{44}
\expandafter\ifx\csname natexlab\endcsname\relax\def\natexlab#1{#1}\fi
\expandafter\ifx\csname bibnamefont\endcsname\relax
  \def\bibnamefont#1{#1}\fi
\expandafter\ifx\csname bibfnamefont\endcsname\relax
  \def\bibfnamefont#1{#1}\fi
\expandafter\ifx\csname citenamefont\endcsname\relax
  \def\citenamefont#1{#1}\fi
\expandafter\ifx\csname url\endcsname\relax
  \def\url#1{\texttt{#1}}\fi
\expandafter\ifx\csname urlprefix\endcsname\relax\def\urlprefix{URL }\fi
\providecommand{\bibinfo}[2]{#2}
\providecommand{\eprint}[2][]{\url{#2}}

\bibitem[{\citenamefont{{Z. Fisk {\em et al.}}}(1979)}]{Fisk:1979}
\bibinfo{author}{\bibnamefont{{Z. Fisk {\em et al.}}}}, \bibinfo{journal}{J.
  Appl. Phys.} \textbf{\bibinfo{volume}{50}}, \bibinfo{pages}{1911}
  (\bibinfo{year}{1979}).

\bibitem[{\citenamefont{{L. Degiorgi {\em et al.}}}(1997)}]{Degiorgi:1997}
\bibinfo{author}{\bibnamefont{{L. Degiorgi {\em et al.}}}},
  \bibinfo{journal}{Phys. Rev. Lett.} \textbf{\bibinfo{volume}{79}},
  \bibinfo{pages}{5134} (\bibinfo{year}{1997}).

\bibitem[{\citenamefont{{S. S{\"u}llow {\em et al.}}}(1998)}]{Sullow:1998}
\bibinfo{author}{\bibnamefont{{S. S{\"u}llow {\em et al.}}}},
  \bibinfo{journal}{Phys. Rev. B} \textbf{\bibinfo{volume}{57}},
  \bibinfo{pages}{5860} (\bibinfo{year}{1998}).

\bibitem[{\citenamefont{{S. Paschen {\em et al.}}}(2000)}]{Paschen:2000}
\bibinfo{author}{\bibnamefont{{S. Paschen {\em et al.}}}},
  \bibinfo{journal}{Phys. Rev. B} \textbf{\bibinfo{volume}{61}},
  \bibinfo{pages}{4174} (\bibinfo{year}{2000}).

\bibitem[{\citenamefont{{W. Henggeler {\em et al.}}}(1998)}]{Henggeler:1998}
\bibinfo{author}{\bibnamefont{{W. Henggeler {\em et al.}}}},
  \bibinfo{journal}{Sol. State Comm.} \textbf{\bibinfo{volume}{108}},
  \bibinfo{pages}{929} (\bibinfo{year}{1998}).

\bibitem[{\citenamefont{{G. A. Wigger {\em et al.}}}(2002)}]{Wigger:2002}
\bibinfo{author}{\bibnamefont{{G. A. Wigger {\em et al.}}}},
  \bibinfo{journal}{Phys. Rev. B} \textbf{\bibinfo{volume}{66}},
  \bibinfo{pages}{212410} (\bibinfo{year}{2002}).

\bibitem[{\citenamefont{{S. Broderick {\em et
  al.}}}(2002{\natexlab{a}})}]{Broderick:2002}
\bibinfo{author}{\bibnamefont{{S. Broderick {\em et al.}}}},
  \bibinfo{journal}{Phys. Rev. B} \textbf{\bibinfo{volume}{65}},
  \bibinfo{pages}{121102(R)} (\bibinfo{year}{2002}{\natexlab{a}}).

\bibitem[{\citenamefont{{S. Broderick {\em et al.}}}(2003)}]{Broderick:2003}
\bibinfo{author}{\bibnamefont{{S. Broderick {\em et al.}}}},
  \bibinfo{journal}{Eur. Phys. J. B} \textbf{\bibinfo{volume}{33}},
  \bibinfo{pages}{47} (\bibinfo{year}{2003}).

\bibitem[{\citenamefont{{A. Perucchi {\em et al.}}}(2003)}]{Perucchi:2003}
\bibinfo{author}{\bibnamefont{{A. Perucchi {\em et al.}}}},
  \bibinfo{journal}{Phys. Rev. Lett.} \textbf{\bibinfo{volume}{92}},
  \bibinfo{pages}{67401} (\bibinfo{year}{2003}).

\bibitem[{\citenamefont{{G. Caimi {\em et al.}}}(2003)}]{Caimi:2004}
\bibinfo{author}{\bibnamefont{{G. Caimi {\em et al.}}}},
  \bibinfo{journal}{Phys. Rev. B} \textbf{\bibinfo{volume}{69}},
  \bibinfo{pages}{12406} (\bibinfo{year}{2003}).

\bibitem[{\citenamefont{{G.A. Wigger {\em et al.}}}(2003)}]{Wigger:2003}
\bibinfo{author}{\bibnamefont{{G.A. Wigger {\em et al.}}}},
  \bibinfo{journal}{cond-mat/0309412}  (\bibinfo{year}{2003}).

\bibitem[{\citenamefont{{S. Massidda {\em et al.}}}(1996)}]{Massidda:1996}
\bibinfo{author}{\bibnamefont{{S. Massidda {\em et al.}}}},
  \bibinfo{journal}{Z. Phys. B.} \textbf{\bibinfo{volume}{102}},
  \bibinfo{pages}{83} (\bibinfo{year}{1996}).

\bibitem[{\citenamefont{{C. O. Rodriguez {\em et al.}}}(2000)}]{Rodriguez:2000}
\bibinfo{author}{\bibnamefont{{C. O. Rodriguez {\em et al.}}}},
  \bibinfo{journal}{Phys. Rev. Lett.} \textbf{\bibinfo{volume}{84}},
  \bibinfo{pages}{3903} (\bibinfo{year}{2000}).

\bibitem[{\citenamefont{{M. C. Aronson {\em et al.}}}(1999)}]{Aronson:1999}
\bibinfo{author}{\bibnamefont{{M. C. Aronson {\em et al.}}}},
  \bibinfo{journal}{Phys. Rev. B} \textbf{\bibinfo{volume}{59}},
  \bibinfo{pages}{4720} (\bibinfo{year}{1999}).

\bibitem[{\citenamefont{{It should be stressed that this experiments require
  large magnetic fields.}}()}]{Foot1}
\bibinfo{author}{\bibnamefont{{It should be stressed that this experiments
  require large magnetic fields.}}}

\bibitem[{\citenamefont{{J. D. Denlinger {\em et al.}}}(2002)}]{Denlinger:2002}
\bibinfo{author}{\bibnamefont{{J. D. Denlinger {\em et al.}}}},
  \bibinfo{journal}{Phys. Rev. Lett.} \textbf{\bibinfo{volume}{89}},
  \bibinfo{pages}{157601} (\bibinfo{year}{2002}).

\bibitem[{\citenamefont{{S. Souma {\em et al.}}}(2003)}]{Souma:2003}
\bibinfo{author}{\bibnamefont{{S. Souma {\em et al.}}}},
  \bibinfo{journal}{Phys. Rev. Lett.} \textbf{\bibinfo{volume}{90}},
  \bibinfo{pages}{27202} (\bibinfo{year}{2003}).

\bibitem[{\citenamefont{{J. C. Cooley {\em et al.}}}(1997)}]{Cooley:1997}
\bibinfo{author}{\bibnamefont{{J. C. Cooley {\em et al.}}}},
  \bibinfo{journal}{Phys. Rev. B} \textbf{\bibinfo{volume}{56}},
  \bibinfo{pages}{14541} (\bibinfo{year}{1997}).

\bibitem[{\citenamefont{{M.J. Calder{\'o}n {\em et
  al.}}}(2003)}]{Calderon:2003}
\bibinfo{author}{\bibnamefont{{M.J. Calder{\'o}n {\em et al.}}}},
  \bibinfo{journal}{cond-mat/0312437}  (\bibinfo{year}{2003}).

\bibitem[{\citenamefont{{M. A. Noack {\em et al.}}}(1980)}]{Noack:1980}
\bibinfo{author}{\bibnamefont{{M. A. Noack {\em et al.}}}},
  \bibinfo{journal}{J. of Cryst. Growth} \textbf{\bibinfo{volume}{49}},
  \bibinfo{pages}{595} (\bibinfo{year}{1980}).

\bibitem[{\citenamefont{{R. Monnier {\em et al.}}}(2001)}]{Monnier:2001}
\bibinfo{author}{\bibnamefont{{R. Monnier {\em et al.}}}},
  \bibinfo{journal}{Phys. Rev. Lett.} \textbf{\bibinfo{volume}{87}},
  \bibinfo{pages}{157204} (\bibinfo{year}{2001}).

\bibitem[{\citenamefont{{V. M. Pereira {\em et al}}}(2003)}]{Pereira:2003}
\bibinfo{author}{\bibnamefont{{V. M. Pereira {\em et al}}}},
  \bibinfo{journal}{unpublished}  (\bibinfo{year}{2003}).

\bibitem[{\citenamefont{Anderson and Hasegawa}(1955)}]{Anderson:1955}
\bibinfo{author}{\bibfnamefont{P.~W.} \bibnamefont{Anderson}} \bibnamefont{and}
  \bibinfo{author}{\bibfnamefont{H.}~\bibnamefont{Hasegawa}},
  \bibinfo{journal}{Phys. Rev.} \textbf{\bibinfo{volume}{100}},
  \bibinfo{pages}{675} (\bibinfo{year}{1955}).

\bibitem[{\citenamefont{{The is a valid approximation in the large--$S$
  limit.}}()}]{Foot2}
\bibinfo{author}{\bibnamefont{{The is a valid approximation in the large--$S$
  limit.}}}

\bibitem[{\citenamefont{Dagotto}(2003)}]{Dagotto_Book:2003}
\bibinfo{author}{\bibfnamefont{E.}~\bibnamefont{Dagotto}},
  \emph{\bibinfo{title}{Nanoscale Phase Separation and Colossal
  Magnetoresistance}} (\bibinfo{publisher}{Springer Verlag},
  \bibinfo{year}{2003}).

\bibitem[{\citenamefont{Economou and Cohen}(1972)}]{Economou:1972}
\bibinfo{author}{\bibfnamefont{E.~N.} \bibnamefont{Economou}} \bibnamefont{and}
  \bibinfo{author}{\bibfnamefont{M.~H.} \bibnamefont{Cohen}},
  \bibinfo{journal}{Phys. Rev. B} \textbf{\bibinfo{volume}{5}},
  \bibinfo{pages}{2931} (\bibinfo{year}{1972}).

\bibitem[{\citenamefont{Varma}(1996)}]{Varma:1996}
\bibinfo{author}{\bibfnamefont{C.~M.} \bibnamefont{Varma}},
  \bibinfo{journal}{Phys. Rev. B} \textbf{\bibinfo{volume}{54}},
  \bibinfo{pages}{7328} (\bibinfo{year}{1996}).

\bibitem[{\citenamefont{{Q. Li {\em et al.}}}(1997)}]{Li:1997}
\bibinfo{author}{\bibnamefont{{Q. Li {\em et al.}}}}, \bibinfo{journal}{Phys.
  Rev. B} \textbf{\bibinfo{volume}{56}}, \bibinfo{pages}{4541}
  (\bibinfo{year}{1997}).

\bibitem[{\citenamefont{Haydock}(1980)}]{Haydock:1980}
\bibinfo{author}{\bibfnamefont{R.}~\bibnamefont{Haydock}}, in
  \emph{\bibinfo{booktitle}{Solid State Physics}} (\bibinfo{publisher}{Academic
  Press}, \bibinfo{address}{New York}, \bibinfo{year}{1980}),
  vol.~\bibinfo{volume}{35}, pp. \bibinfo{pages}{216--294}.

\bibitem[{\citenamefont{MacKinnon and Kramer}(1981)}]{MacKinnon:1981}
\bibinfo{author}{\bibfnamefont{A.}~\bibnamefont{MacKinnon}} \bibnamefont{and}
  \bibinfo{author}{\bibfnamefont{B.}~\bibnamefont{Kramer}},
  \bibinfo{journal}{Phys. Rev. Lett.} \textbf{\bibinfo{volume}{47}},
  \bibinfo{pages}{1546} (\bibinfo{year}{1981}).

\bibitem[{\citenamefont{Pichard and Sarma}(1981)}]{Pichard:1981}
\bibinfo{author}{\bibfnamefont{J.~L.} \bibnamefont{Pichard}} \bibnamefont{and}
  \bibinfo{author}{\bibfnamefont{G.}~\bibnamefont{Sarma}}, \bibinfo{journal}{J.
  Phys. C} \textbf{\bibinfo{volume}{14}}, \bibinfo{pages}{L127}
  (\bibinfo{year}{1981}).

\bibitem[{\citenamefont{Maldague}(1977)}]{Maldague:1977}
\bibinfo{author}{\bibfnamefont{P.~F.} \bibnamefont{Maldague}},
  \bibinfo{journal}{Phys. Rev. B} \textbf{\bibinfo{volume}{16}},
  \bibinfo{pages}{2737} (\bibinfo{year}{1977}).

\bibitem[{\citenamefont{Wooten}(1972)}]{Wooten:1972}
\bibinfo{author}{\bibfnamefont{F.}~\bibnamefont{Wooten}},
  \emph{\bibinfo{title}{Optical Properties of Solids}}
  (\bibinfo{publisher}{Academic Press}, \bibinfo{address}{New York},
  \bibinfo{year}{1972}).

\bibitem[{\citenamefont{{S. Broderick {\em et
  al.}}}(2002{\natexlab{b}})}]{Broderick:2002b}
\bibinfo{author}{\bibnamefont{{S. Broderick {\em et al.}}}},
  \bibinfo{journal}{Eur. Phys. J. B} \textbf{\bibinfo{volume}{27}},
  \bibinfo{pages}{3} (\bibinfo{year}{2002}{\natexlab{b}}).

\bibitem[{\citenamefont{Lee and Ramakrishnan}(1985)}]{Lee:1985}
\bibinfo{author}{\bibfnamefont{P.~A.} \bibnamefont{Lee}} \bibnamefont{and}
  \bibinfo{author}{\bibfnamefont{T.~V.} \bibnamefont{Ramakrishnan}},
  \bibinfo{journal}{Rev. Mod. Phys.} \textbf{\bibinfo{volume}{57}},
  \bibinfo{pages}{287} (\bibinfo{year}{1985}).

\bibitem[{\citenamefont{de~Gennes}(1960)}]{Gennes:1960}
\bibinfo{author}{\bibfnamefont{P.-G.} \bibnamefont{de~Gennes}},
  \bibinfo{journal}{Phys. Rev.} \textbf{\bibinfo{volume}{118}},
  \bibinfo{pages}{141} (\bibinfo{year}{1960}).

\bibitem[{\citenamefont{Kubo and Ohata}(1972)}]{Kubo:1972}
\bibinfo{author}{\bibfnamefont{K.}~\bibnamefont{Kubo}} \bibnamefont{and}
  \bibinfo{author}{\bibfnamefont{N.}~\bibnamefont{Ohata}}, \bibinfo{journal}{J.
  Phys. Soc. Jpn.} \textbf{\bibinfo{volume}{33}}, \bibinfo{pages}{21}
  (\bibinfo{year}{1972}).

\bibitem[{\citenamefont{{Yonathan Shapir {\em et al.}}}(1982)}]{Shapir:1982}
\bibinfo{author}{\bibnamefont{{Yonathan Shapir {\em et al.}}}},
  \bibinfo{journal}{Phys. Rev. Lett.} \textbf{\bibinfo{volume}{49}},
  \bibinfo{pages}{486} (\bibinfo{year}{1982}).

\bibitem[{\citenamefont{Mott and Davis}(1971)}]{Mott:1971}
\bibinfo{author}{\bibfnamefont{N.~F.} \bibnamefont{Mott}} \bibnamefont{and}
  \bibinfo{author}{\bibfnamefont{E.~A.} \bibnamefont{Davis}},
  \emph{\bibinfo{title}{Electronic processes in non-crystalline materials}}
  (\bibinfo{publisher}{Clarendon Press}, \bibinfo{address}{Oxford},
  \bibinfo{year}{1971}).

\bibitem[{\citenamefont{{I. Chang {\em et al.}}}(1994)}]{Chang:1994}
\bibinfo{author}{\bibnamefont{{I. Chang {\em et al.}}}},
  \bibinfo{journal}{Phys. Rev. Lett.} \textbf{\bibinfo{volume}{74}},
  \bibinfo{pages}{2094} (\bibinfo{year}{1994}).

\bibitem[{\citenamefont{Stauffer and Aharony}(1994)}]{Stauffer:1994}
\bibinfo{author}{\bibfnamefont{D.}~\bibnamefont{Stauffer}} \bibnamefont{and}
  \bibinfo{author}{\bibfnamefont{A.}~\bibnamefont{Aharony}},
  \emph{\bibinfo{title}{{Introduction to Percolation Theory}}}
  (\bibinfo{publisher}{Taylor \& Francis}, \bibinfo{address}{London},
  \bibinfo{year}{1994}), \bibinfo{edition}{2nd} ed.

\bibitem[{\citenamefont{Wegener and Littlewood}(2002)}]{Wegener:2002}
\bibinfo{author}{\bibfnamefont{L.~G.~L.} \bibnamefont{Wegener}}
  \bibnamefont{and} \bibinfo{author}{\bibfnamefont{P.~B.}
  \bibnamefont{Littlewood}}, \bibinfo{journal}{Phys. Rev. B}
  \textbf{\bibinfo{volume}{66}}, \bibinfo{pages}{224402}
  (\bibinfo{year}{2002}).

\bibitem[{\citenamefont{{P. Nyhus {\em et al.}}}(1997)}]{Nyhus:1997}
\bibinfo{author}{\bibnamefont{{P. Nyhus {\em et al.}}}},
  \bibinfo{journal}{Phys. Rev. B} \textbf{\bibinfo{volume}{56}},
  \bibinfo{pages}{2717} (\bibinfo{year}{1997}).

\bibitem[{\citenamefont{{C. S. Snow {\em et al.}}}(2001)}]{Snow:2001}
\bibinfo{author}{\bibnamefont{{C. S. Snow {\em et al.}}}},
  \bibinfo{journal}{Phys. Rev. B} \textbf{\bibinfo{volume}{64}},
  \bibinfo{pages}{174412} (\bibinfo{year}{2001}).

\end{thebibliography}

\end{document}